\documentclass[10pt]{article}
\usepackage{euscript,amssymb,amsmath}
\usepackage[dvips]{graphicx}
\catcode`\@=11

\newcommand{\be}[1]{\begin{equation}\label{#1}}
\newcommand{\ee}{\end{equation}}
\newcommand{\ba}[1]{\begin{eqnarray}\label{#1}}
\newcommand{\ea}{\end{eqnarray}}
\newcommand{\rf}[1]{(\ref{#1})}
\newcommand{\nn}{\nonumber}

\newcommand{\diag}{\mbox{\rm diag}\,}

\title{Unfolding the conical zones of the dissipation-induced subcritical flutter for the
rotationally symmetrical gyroscopic systems
 \\ {\large Oleg N. Kirillov}\footnote{Dynamics group, Department of Mechanical Engineering,
 Technical University of Darmstadt,
 Hochschulstr. 1, 64289 Darmstadt, Germany
 (e-mail:~kirillov@dyn.tu-darmstadt.de, Tel: +49 6151 16 6828, Fax: +49 6151 16 4125) and
Institute of Mechanics, Moscow State Lomonosov University, Michurinskii pr. 1, 119192 Moscow,
Russia (e-mail:~kirillov@imec.msu.ru).}}
\date{}

\begin{document}
\maketitle
\begin{abstract}
Flutter of an elastic body of revolution spinning about its axis of symmetry is prohibited in the subcritical spinning speed range by the Krein theorem for the Hamiltonian perturbations. Indefinite damping creates conical domains of the subcritical flutter (subcritical parametric resonance) bifurcating into the pockets of two Whitney's umbrellas when non-conservative positional forces are additionally taken into account. This explains why in contrast to the common intuition, but in agreement with experience,
symmetry-breaking stiffness variation can promote subcritical friction-induced oscillations of the rotor rather than inhibit them.
\end{abstract}

\begin{flushleft}
{\small PACS: 45.30.+s, 46.40.Ff, 05.45.Xt, 45.20.dc, 45.10.Hj, 02.10.Yn, 02.40.Xx
Submitted to: {\it Phys. Lett. A}}
\end{flushleft}

\section{Introduction}

Consider an autonomous linear gyroscopic system describing small oscillations in the discretized models of rotating elastic bodies of revolution considered in a stationary frame \cite{H78,Ve95,NN98,HL01,SHKH07,Ki08b}
\be{i1}
\ddot{\bf x} + 2\Omega{\bf G}\dot{\bf x} + ({\bf P}+\Omega^2{\bf G}^2) {\bf x}=0, \quad {\bf x}=\mathbb{R}^{2n},
\ee
where $\Omega$ is the speed of rotation, ${\bf P}$=${\rm diag}(\omega_1^2,\omega_1^2,\omega_2^2,\omega_2^2,\ldots,\omega_n^2,\omega_n^2)={\bf P}^T$ is the matrix of potential forces, and ${\bf G}$=${\rm blockdiag}({\bf J},2{\bf J},\ldots,n{\bf J})=-{\bf G}^T$ is the matrix of gyroscopic forces with
\be{i2}
{\bf J}=\left(
                                                                                      \begin{array}{cc}
                                                                                        0 & -1 \\
                                                                                        1 & 0 \\
                                                                                      \end{array}
                                                                                    \right).
\ee
Due to rotational symmetry of the rotor and periodic boundary conditions the eigenvalues $0<\omega_1^2<\omega_2^2<\cdots<\omega_{n-1}^2<\omega_n^2$ of the matrix $\bf P$ are double semi-simple, that is each eigenvalue
$\omega_s^2$ has two linearly independent eigenvectors \cite{KMS05a,KMS05b}.
The distribution of the doublets $\omega_s^2$ as a function of $s$ is usually different for various bodies of revolution.
For example, $\omega_s=s$ corresponds to the spectrum of a circular string \cite{Ki08b}.
Nevertheless, there can exist isospectral bodies, because "one cannot hear the shape of a drum" \cite{GWW92}.

Separating time by the substitution ${\bf x}={\bf u}\exp(\lambda t)$, we arrive at the eigenvalue problem for the operator ${\bf L}_0$
\be{i3}
{\bf L}_0(\Omega){\bf u}:=({\bf I}\lambda^2 + 2\Omega{\bf G}\lambda + {\bf P}+\Omega^2{\bf G}^2) {\bf u}=0.
\ee
As a consequence of the block-diagonal structure of the sparse matrices $\bf G$ and $\bf P$, the eigenvalues of ${\bf L}_0$ are
\be{i4}
\lambda_s^{\pm}=i\omega_s\pm is\Omega,\quad \overline{\lambda_s^{\pm}}=-i\omega_s\mp is\Omega,
\ee
where bar over a symbol denotes complex conjugate. Rotation causes the doublet modes $\pm i\omega_s$ to split.
The newborn pair of simple eigenvalues $\lambda_s^{\pm}$ corresponds to the forward and backward traveling waves,
which propagate along the circumferential direction.
Viewed from the stationary frame, the frequency of the forward traveling wave appears to
increase and that of the backward traveling wave appears to decrease, as the spin increases.
Double eigenvalues thus originate again at non-zero angular velocities,
forming the nodes of the \textit{spectral mesh} \cite{GK06} of the crossed frequency curves
in the plane \lq frequency' versus \lq angular velocity'.
The spectral meshes generically appear in equivariant dynamical systems \cite{DM94} and are characteristic for such rotating symmetric continua as circular strings, discs, rings and cylindrical and hemispherical shells,
vortex rings, and a spherically symmetric $\alpha^2$-dynamo
of magnetohydrodynamics \cite{Ki08b,GK06}.

At the angular velocity $\Omega_s^{cr}=\omega_s/s$
the frequency of the $s$th backward traveling wave vanishes to zero $(\lambda_s^{\pm}=\overline{\lambda_s^{\pm}}=0)$, so that the wave remains stationary in the non-rotating
frame. The lowest one of such velocities, $\Omega_{cr}$, is called \textit{critical} \cite{Mo98}. When the speed of rotation exceeds the critical speed, some backward waves, corresponding to
the eigenvalues $\overline{\lambda_s^{\pm}}$,
travel slower than the disc rotation speed and appear to be traveling forward (reflected waves).
The effective energy of the reflected wave is negative and that of the forward and backward traveling
waves is positive \cite{MS86}. Therefore, in the \textit{subcritical} speed region $|\Omega|<\Omega_{cr}$ all the crossings of the
frequency curves correspond to the forward and backward modes of the same signature,
while in the \textit{supercritical} speed region $|\Omega|>\Omega_{cr}$ there exist crossings that are formed by the reflected and forward/backward modes of opposite signature. As a consequence of Krein's theorem \cite{MS86,HM87}, under Hamiltonian perturbations
the crossings in the subcritical region veer away into \textit{avoided crossings} (stability), while in the supercritical region the crossings with the mixed signature turn into the rings of complex
eigenvalues---\textit{bubbles of instability} \cite{MS86}---leading to flutter known also
as the \lq mass and stiffness instabilities' \cite{Ki08b}.

A \textit{supercritical flutter} frequently occurs in the high speed applications like circular saws and computer storage devices,
while the \textit{subcritical flutter}---either desirable as a source of instability at low speeds as in the case of musical instruments like the singing wine glass and a glass harmonica or undesirable as in the case of the squealing disc- and drum brakes \cite{SHKH07,Ki08b,Mo98,Sp61}---is an elusive phenomenon. The latter property is characteristic for the regions with the definite Krein signature and appears, for example, as the subtleties in the tuning the helical turbulence parameter for excitation of the oscillatory MHD dynamo \cite{KGS08}.

Being prohibited by Krein's theorem for the Hamiltonian systems, subcritical flutter can occur, however, due to non-Hamiltonian, i.e. dissipative and non-conservative, perturbations \cite{Ki08b,Br97}. Since even the simplest
codimension-1 Hamiltonian Hopf bifurcation can be viewed as a singular limit of the codimension-3
dissipative resonant $1:1$ normal form and the essential singularity in which these two cases meet is topologically equivalent to
Whitney's umbrella \cite{HR95,Ki06,KM07}, substantially complicating the stability analysis \cite{MK91,Ki07,SBM08},
one can expect that exactly the singularities, associated with the double eigenvalues of the spectral mesh, make the
induction of subcritical flutter in system \rf{i1} by non-Hamiltonian perturbations a challenging problem.

In this Letter using perturbation theory of multiple eigenvalues we show that even if the eigenvalue
branches in the subcritical region are well separated at the avoided crossings, created by the stiffness variation,
they can be forced to bend with the origination of arcs of complex eigenvalues with positive real parts by the indefinite damping,
which comes to the equations of motion, e.g., from the negative friction-velocity gradient \cite{Sp61,KNP08,KP08}.
In the space of the gyroscopic, damping, and stiffness parameters the zones of the subcritical
dissipation-induced flutter turn out to be cones with the apexes at the points, corresponding to the nodes of the spectral mesh.
We show that the orientation of the cone is substantially determined by the structure of the damping matrix, which can be chosen
in such a way that the system is unstable for significant magnitudes of the stiffness matrix detuning. The conical zones of
the subcritical flutter bifurcate into couples of Whitney's umbrellas when non-conservative positional forces
(originated from the follower forces or from the moment generated by variation in friction forces \cite{Mo98}) are additionally taken into account. We describe in detail this process related to the unfolding of the conical eigenvalue surface near a diabolical point to the \lq \lq double coffee filter" singularity with two exceptional points under a complex perturbation of a real symmetric matrix in the problems of wave propagation in chiral and dissipative media \cite{KMS05a,KMS05b,K75,KKM03,BD03}.

\section{Conical zones of the dissipation-induced subcritical flutter}

The models of disc brakes, where subcritical flutter was detected by numerical approaches, include both the case when the point-wise or distributed friction pads are rotated
around a stationary disc, affecting a point or a sector of it, and when the disc rotates past the stationary friction pads, see \cite{Ki08b,Mo98} and references therein. Linearization and discretization of the latter class of the models frequently results in the
equation \rf{i1} perturbed by the matrices ${\bf D}=(d_{ij})={\bf D}^T$, ${\bf K}=(k_{ij})={\bf K}^T$, and ${\bf N}=(n_{ij})=-{\bf N}^T$
of dissipative, potential, and non-conservative positional forces, respectively:
\be{i5}
\ddot{\bf x} + (2\Omega{\bf G}+\delta{\bf D})\dot{\bf x} + ({\bf P}+\Omega^2{\bf G}^2+\kappa {\bf K} + \nu {\bf N}) {\bf x}=0,
\ee
where the parameters $\delta$, $\kappa$, and $\nu$ control the magnitudes of the perturbations. The matrices $\bf D$, $\bf K$, and $\bf N$ can be assumed to be functions of $\Omega$.

In the rotating frame the load appears to be moving periodically in the circumferential direction.
The transformation ${\bf x}={\bf A}{\bf z}:=\exp(-\Omega{\bf G}t){\bf z}$, see \cite{K01,Sh01}, yields an equivalent to \rf{i5} potential system with the time-periodic perturbation of the coefficients
\be{i6}
\ddot{\bf z}+\delta\widetilde{\bf D}(t)\dot{\bf z}+({\bf P}-\delta\Omega\widetilde{\bf D}(t)\widetilde{\bf G}(t)+\kappa\widetilde{\bf K}(t)+\nu\widetilde{\bf N}(t)){\bf z}=0,
\ee
where $\widetilde{\bf D}(t)={\bf A}^{-1}{\bf D}{\bf A}$, $\widetilde{\bf G}(t)={\bf A}^{-1}{\bf G}{\bf A}={\bf G}$,
$\widetilde{\bf K}(t)={\bf A}^{-1}{\bf K}{\bf A}$, $\widetilde{\bf N}(t)={\bf A}^{-1}{\bf N}{\bf A}$.
For $2n=2$ degrees of freedom the matrix $\widetilde{\bf N}(t)={\bf N}$, while the periodic stiffness and damping matrices
have a simple explicit form
\ba{i7}
2\widetilde{\bf K}(t)&=&\diag({\rm tr}{\bf K},{\rm tr}{\bf K}) +({\bf K}+{\bf J}{\bf K}{\bf J})\cos(2\Omega t)+({\bf J}{\bf K}-{\bf K}{\bf J})\sin(2\Omega t),\nn \\
2\widetilde{\bf D}(t)&=&\diag({\rm tr}{\bf D},{\rm tr}{\bf D}) +({\bf D}+{\bf J}{\bf D}{\bf J})\cos(2\Omega t)+({\bf J}{\bf D}-{\bf D}{\bf J})\sin(2\Omega t).
\ea
Note that in recent Letters \cite{Sh01,Sh04}, Shapiro studied only a particular case of system \rf{i6}, \rf{i7} with $2n=2$ degrees of freedom,
${\bf D}=\diag(1,1)$, ${\bf K}=\diag(-1,1)$, ${\bf N}={\bf J}$, $\omega_1^2=a$, $\kappa=2q$, $\delta=2h$, and $\nu=2h\Omega$, which is, however, used much in modeling strong focusing in electron and ion optics through a twisted series of static
quadrupole magnetic or electric lenses or helical quadrupole channels as well as in modeling the monochromatic light of two
orthogonal polarizations propagating along the axis of helically twisted anisotropic media like cholesteric liquid crystals, see
\cite{Sh01,Sh04} and references therein.

    \begin{figure}
    \begin{center}
    \includegraphics[angle=0, width=0.99\textwidth]{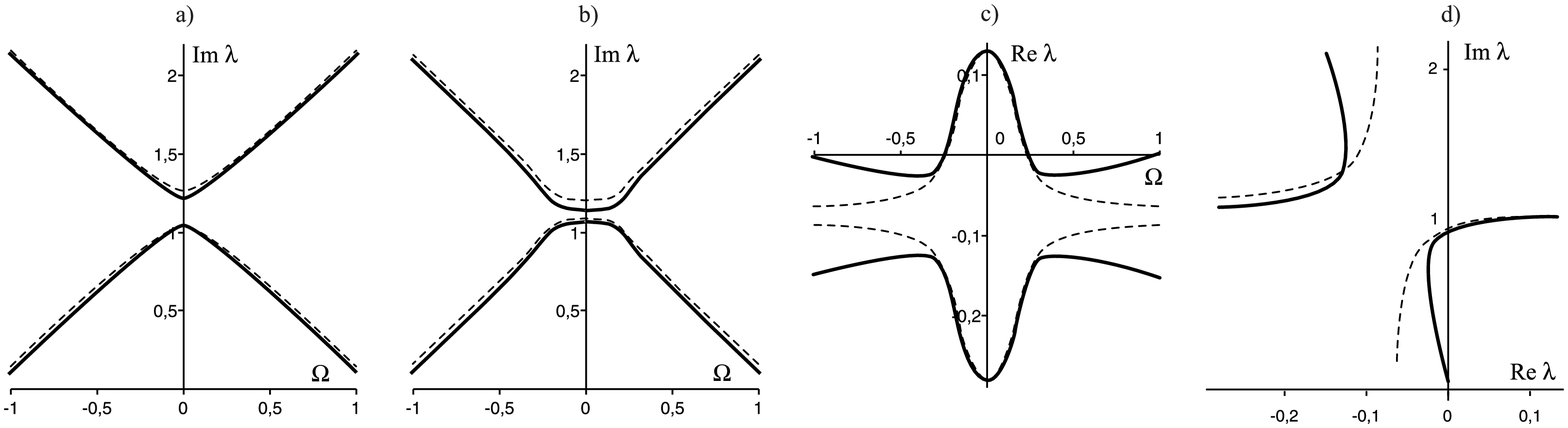}
    \end{center}
    \caption{Numerically calculated eigenvalue branches (thick lines) and their approximations \rf{t5} (dashed lines) for $\omega_1=1$,
    $k_{11}=1$, $k_{22}=2$, $k_{12}=1$, $d_{11}=-1$, $d_{22}=2$, $d_{12}=0$, $\nu=0$, $\kappa=0.2$, and (a) $\delta=0$, (b)-(d) $\delta=0.3$.  }
    \label{fig1}
    \end{figure}

In the absence
of dissipative $(\delta=0)$ and non-conservative positional $(\nu=0)$ terms, \rf{i6} describes coupled parametric oscillators
of Mathieu-type with
the periodic in time potential possessing parametric resonance in the supercritcal range ~$|\Omega|>\Omega_{cr}$.
Inclusion of parametrically excited damping and non-conservative terms makes the equation \rf{i6} a less traditional
periodic system demonstrating parametric resonance in both the subcritical and supercritical regions \cite{Mo98}.

Nevertheless, the equivalence of the two dual descriptions enables us to reduce the investigation of the periodic system \rf{i6} to a considerably simpler study of the stability of the autonomous system \rf{i5}, cf. \cite{K01,Sh01}. In particular, subcritical parametric resonance domains of equation \rf{i6} correspond to the regions of subcritical flutter of the system \rf{i5}.

Owing to its relative simplicity the case of two degrees of freedom $(n=1)$ allows for the detailed stability analysis.
For $n=1$ the spectrum of the unperturbed operator ${\bf L}_0(\Omega)$ consists of four branches \rf{i4}, forming the
spectral mesh in the $(\Omega,{\rm Im}\lambda)$-plane.
In the subcritical region $|\Omega|<\Omega_{cr}=\omega_1$ the branches cross
at the points $(0,\pm \omega_1)$. Assuming without loss in generality ${\bf N}={\bf J}$, we consider a general perturbation of the gyroscopic system ${\bf L}_0(\Omega)+\Delta {\bf L}(\Omega)$. The size of the perturbation $\Delta {\bf
L}(\Omega)=\delta\lambda{\bf D}+\kappa{\bf K}+\nu {\bf N}\sim
\varepsilon$ is small, where $\varepsilon=\| \Delta{\bf L}(0) \|$ is
the Frobenius norm of the perturbation at $\Omega=0$.
For small $\Omega$ and $\varepsilon$ perturbation of the double semi-simple eigenvalue $\lambda=i\omega_1$ with
two orthogonal eigenvectors ${\bf u}_1=(0,(2\omega_1)^{-1/2})$ and ${\bf u}_2=((2\omega_1)^{-1/2},0)$
is described by the formulas \cite{Ki08b}
\be{t5}
{\rm Re}\lambda=-\frac{\mu_1+\mu_2}{4}\delta\pm
\sqrt{\frac{|c|+{\rm Re}c}{2}},\qquad
{\rm Im}\lambda=\omega_1+\frac{\rho_1+\rho_2}{4\omega_1}\kappa\pm
\sqrt{\frac{|c|-{\rm Re}c}{2}},
\ee
\be{t6}
{\rm Re}c=\left(\frac{\mu_1{-}\mu_2}{4}\right)^2\delta^2-\left(\frac{\rho_1{-}\rho_2}{4\omega_1} \right)^2\kappa^2-
\Omega^2+\frac{\nu^2}{4\omega_1^2},\qquad
{\rm Im}c=\frac{\Omega\nu}{\omega_1}-\delta\kappa\frac{2{\rm tr}{\bf KD}-
{\rm tr}{\bf K}{\rm tr}{\bf D}}{8\omega_1},
\ee
where $\mu_{1,2}$ and $\rho_{1,2}$ are the eigenvalues of the matrices $\bf D$ and $\bf K$, respectively.

    \begin{figure}
    \begin{center}
    \includegraphics[angle=0, width=0.8\textwidth]{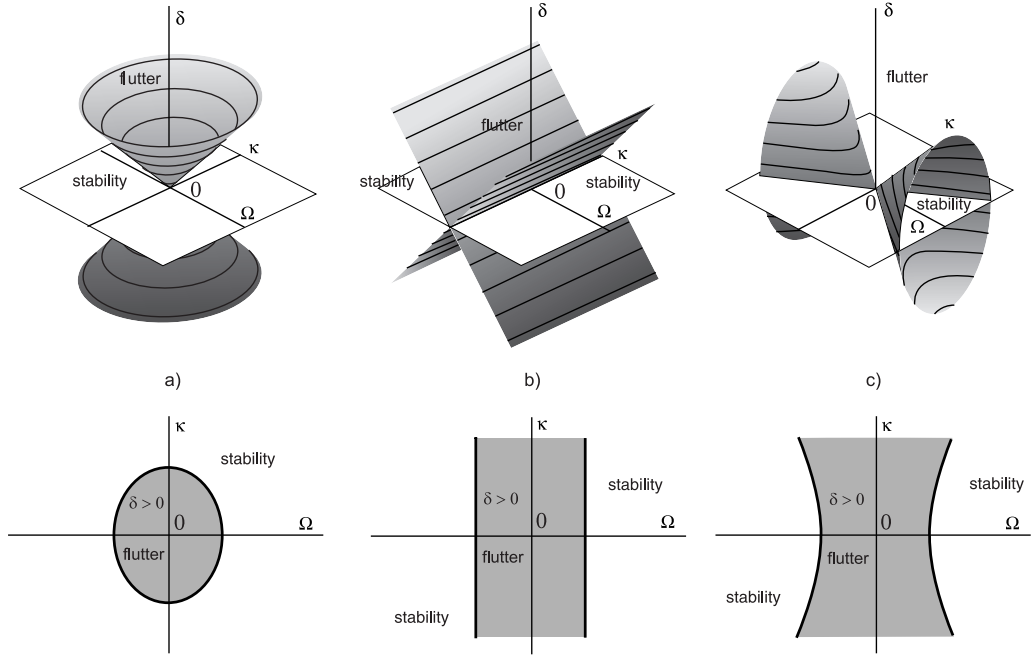}
    \end{center}
    \caption{Domains of the subcritical flutter instability for indefinite damping matrix in the absence
    of the non-conservative positional forces when (a) $A>0$, (b) $A=0$, and (c) $A<0$. }
    \label{fig2}
    \end{figure}

Since the eigenvalues at the crossings in the subcritical range have the same Krein signature, they veer away under potential perturbation $\kappa{\bf K}$, destroying the rotational symmetry of the body, independently on whether the perturbing matrix $\bf K$ is definite or indefinite \cite{Ki08b}.
The detuning of the stiffness matrix ${\bf P}+\kappa{\bf K}$
does not shift the eigenvalues from the imaginary axis, preserving the
marginal stability. Near the node of the spectral mesh $(\Omega=0,{\rm Im}\lambda=\omega_1)$ the veering
is described by the hyperbola \cite{Ki08b}
\be{c1}
\left({\rm Im}\lambda-\omega_1-\frac{\rho_1+\rho_2}{4\omega_1}\kappa\right)^2-\Omega^2=
\left(\frac{\rho_1-\rho_2}{4\omega_1} \right)^2\kappa^2,\quad {\rm Re}\lambda=0,
\ee
shown by the dashed lines in Fig.~\ref{fig1}(a).

Now we demonstarte that even if the eigenvalues of the rotationally symmetric gyroscopic system \rf{i1} are separated
in the subcritical region
by the symmetry-breaking variation of the stiffness matrix $\kappa{\bf K}$, the inclusion of dissipation $\delta{\bf D}$
can cause flutter instability when $\bf D$ is indefinite, see Fig.~\ref{fig1}(b)-(d). Indeed, with $\nu=0$ in equation \rf{t5},
the condition ${\rm Re}\lambda < 0$ yields linear approximation to the domain of asymptotic stability in the $(\Omega, \kappa, \delta)$ - space
\be{cd1}
\delta{\rm tr}{\bf D}>0,\quad \kappa^2A+\Omega^2(2\omega_1 {\rm tr}{\bf D})^2 >-\det{\bf D}(\omega_1 {\rm tr}{\bf D})^2 \delta^2,
\ee
where
\be{cd1a}
A:=\det{\bf D}(\rho_1{-}\rho_2)^2{+}\left(k_{12}(d_{22}{-}d_{11}){-}d_{12}(k_{22}{-}k_{11}) \right)^2
=\frac{(({\rm tr}{\bf D})^2-16\beta_0^2)(\rho_1-\rho_2)^2}{4},
\ee
\be{cd1b}
\beta_0:=\frac{2 {\rm tr}{\bf KD}-{\rm tr}{\bf K} {\rm tr}{\bf D}}{4(\rho_1-\rho_2)}.
\ee

For the definite damping matrices we have $A>0$ because $\det{\bf D}>0$. Consequently, for $\delta{\bf D}>0$ the conditions \rf{cd1}
are always fulfilled in agreement with the Thomson-Tait-Chetaev theorem \cite{Ki07,SBM08}.
Since $\det {\bf D}<0$ in case of indefinite damping, the inequalities \rf{cd1} indicate that the flutter instability domain
for $A>0$ is inside the conical surface extended along the $\delta$-axis and the stability domain is adjacent to the cone's skirt
selected by the condition $\delta{\rm tr}{\bf D}>0$, see Fig.~\ref{fig2}(a).
The conical domain is stretched along the $\kappa$-axis when $A$ tends to zero and it is transformed into a dihedral angle when $A=0$, as shown in Fig.~\ref{fig2}(b). With the further decrease in $A$ the dihedral angle is again wrapped into the conical surface which is then extended along the $\Omega$-axis, Fig.~\ref{fig2}(c). The domain of asymptotic stability is inside the half of the cone selected by the inequality $\delta{\rm tr}{\bf D}>0$. The eigenvalue branches with their approximations \rf{t5}, illustrating
the mechanism of the dissipation-induced subcritical flutter in the presence of imperfections $\kappa{\bf K}$, are shown in
Fig.~\ref{fig1}(b)-(d) for $A<0$. Note that the threshold $A=0$, separating the indefinite damping matrices,
coincides with that found first in \cite{Ki07} from the criteria of Routh and Hurwitz for a general two-dimensional non-conservative gyroscopic system
with dissipation.

In case when $n>1$, the mutual orientation of the conical stability boundaries is expected to be determined also by the Krein signature of the eigenvalues involved into the corresponding crossings at the nodes of the spectral mesh, which is substantially different in the subcritical and in the supercritical regions, similarly to the Krein space induced geometry of the resonance tongues of the spherically-symmetric MHD $\alpha^2$-dynamo \cite{KGS08}.

In the plane $(\Omega,\kappa)$ for a fixed $\delta>0$ the instability domain has, respectively, the form
of an ellipse, Fig.~\ref{fig2}(a), a stripe, Fig.~\ref{fig2}(b), or a region contained between the branches of a hyperbola,
Fig.~\ref{fig2}(c). The latter picture shows that a widely known in the engineering practice approach to the squeal
suppression by the reduction of the rotational symmetry of the rotor is not efficient in the presence of indefinite damping,
which originates from the brake pads with the negative friction-velocity gradient \cite{Sp61,KNP08,KP08}. And daily experience testifies, that an imperfect wine glass and a perforated brake disc still experience friction-induced instabilities and emit sound.

Bifurcation of the stability diagrams with the change of the entries of the matrices $\bf D$ and $\bf K$ can also
explain unsatisfactory reproducibility of experiments with disc brakes \cite{Mo98}. Indeed, some parameters like rotational speed
and pressure on the brake pads can be regulated precisely, while the topography of the pads' surface as well as the material
properties of the pads undergo uncontrollable changes from one run of the experiment to the other owing to the heating,
cooling, and wear \cite{Mo98}. As a consequence, the very same values of stiffness $\kappa$ and angular velocity $\Omega$,
which yield flutter instability and squeal for $A<0$, Fig.~\ref{fig2}(c), make the brake quiet for $A > 0$ owing to
the qualitative change of the stability diagram with the change in the structure of the matrices $\bf D$ and $\bf K$, Fig.~\ref{fig2}(a).
This effect becomes even more pronounced in the presence of non-conservative positional forces, because,
as we demonstrate in the next section, the conical stability boundaries are not structurally stable under small perturbations $\nu {\bf N}$.

    \begin{figure}
    \begin{center}
    \includegraphics[angle=0, width=0.99\textwidth]{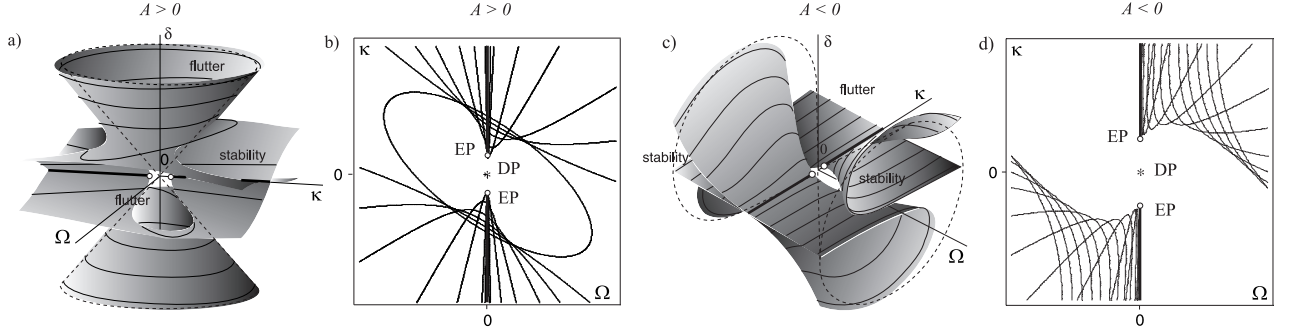}
    \end{center}
    \caption{Unfolding the conical boundary of subcritical flutter (dashed lines) by the perturbation $\nu {\bf N}$ into a couple of Whitney's umbrellas with their level curves depicted for $\delta>0$ when (a),(b) $A>0$ and (c),(d) $A<0$ . }
    \label{fig3}
    \end{figure}

\section{Unfolding the conical zones into couples of Whitney's umbrellas in the presence of non-conservative positional forces}

In the presence of non-conservative positional forces $(\nu \ne 0)$, stability conditions \rf{cd1}, following
from equations \rf{t5} under the requirement ${\rm Re}\lambda < 0$, are modified into $\delta{\rm tr}{\bf D}>0$ and $B>0$, where
\ba{z1}
B:=(2\Omega(\delta^2\omega_1^2{\rm tr}{\bf D}^2-4\nu^2)+\delta(2{\rm tr}{\bf KD}-{\rm tr}{\bf K}{\rm tr}{\bf D})\kappa\nu)^2&+&\delta^2 {\rm tr}{\bf D}^2(A\delta^2\omega_1^2 -\nu^2(\rho_1-\rho_2)^2)\kappa^2 \\ &-&
\delta^2 {\rm tr}{\bf D}^2(\delta^2\omega_1^2{\rm tr}{\bf D}^2-4\nu^2)(\nu^2-\delta^2\omega_1^2\det{\bf D})\nn.
\ea
In the assumption that $\nu=0$ the inequality $B>0$ is reduced to the second one of \rf{cd1}; for $\kappa=0$ it is transformed
into $\Omega^2<\Omega_{cr}^2$, where \cite{Ki08b}
\be{z1a}
\Omega_{cr}=\frac{\delta{\rm tr}{\bf D}}{4}\sqrt{-\frac{\nu^2-\omega_1^2\delta^2\det{\bf D}}{\nu^2-\omega_1^2\delta^2({\rm tr}{\bf D}/2)^2}}.
\ee

For the fixed parameters $\delta$ and $\nu$ the equation $B=0$ generically describes either an ellipse or a hyperbola
in the $(\Omega,\kappa)$-plane, Fig.~\ref{fig3}(b),(d). For $A>0$ and sufficiently big $\delta$, the domain of subcritical
flutter is inside the ellipse $B=0$, and the domain of asymptotic stability is outside the ellipse similarly to the system without
non-conservative positional forces, whose stability diagram is shown in Fig.~\ref{fig2}(a). With the decrease in $\delta$ for the
fixed $\nu$
the ellipse is rotated around the origin in the $(\Omega,\kappa)$-plane and simultaneously it is stretched along one of its main axes.
At the threshold $A\delta^2\omega_1^2 =\nu^2(\rho_1-\rho_2)^2$ the ellipse is transformed into two parallel lines, which with the
further decrease in $\delta$ are bent into two branches of hyperbola $B=0$. Then, the flutter instability domain lies between the
two branches, Fig.~\ref{fig3}(b). When $\delta$ tends to zero, the hyperbolic branches are shrunk
to the union of the intervals $(-\infty,-\kappa_0]\bigcup [\kappa_0,\infty)$ of the $\kappa$-axis, where the critical value
\be{z2}
\kappa_0:=\frac{2\nu}{\rho_1-\rho_2}
\ee
follows from the conditions ${\rm Re}c=0$ and ${\rm Im}c=0$ implying existence of the double eigenvalues at
the points $(\Omega=0,\kappa=\pm \kappa_0,\delta=0)$, which are shown as open circles in  Fig.~\ref{fig3}.

Remarkably, the value \rf{z2} of $\kappa_0$ obtained from the perturbation formulas \rf{t5} coincides with the exact one
following from the characteristic equation of the system \rf{i5} in the assumption that $\delta=0$ and $\Omega=0$
\be{z3}
\lambda^4+(2\omega_1^2+\kappa{\rm tr}{\bf K})\lambda^2+\kappa^2\det{\bf K}+\kappa\omega_1^2{\rm tr}{\bf K}+\nu^2+\omega_1^4=0.
\ee
Substituting $\kappa=\kappa_0$ into \rf{z3} yields the frequency of the corresponding double eigenvalues $\pm i \omega_0$
\be{z4}
\omega_0:=\sqrt{\omega_1^2+\nu \frac{\rho_1+\rho_2}{\rho_1-\rho_2}}=\omega_1+\frac{\nu}{2\omega_1}\frac{\rho_1+\rho_2}{\rho_1-\rho_2} +o(\nu),
\ee
where the first-order approximation with respect to $\nu$ follows also from the formulas \rf{t5} and \rf{z2}.

The double eigenvalue $i\omega_0$ has the eigenvector ${\bf u}_0$ and the associated vector ${\bf u}_1$ of the Jordan chain
\be{z5}
{\bf u}_0=\left(
            \begin{array}{c}
              k_{11}-k_{22} \\
              2k_{12}+\rho_1-\rho_2 \\
            \end{array}
          \right),\quad
{\bf u}_1=-\frac{2i\omega_0(\rho_1-\rho_2)}{\nu}\left(
            \begin{array}{c}
              1 \\
              0 \\
            \end{array}
          \right),
\ee
which are the solutions to the equations
\be{z6}
(-\omega_0^2{\bf I}+{\bf P}+\kappa_0{\bf K}){\bf u}_0=0,\quad (-\omega_0^2{\bf I}+{\bf P}+\kappa_0{\bf K}){\bf u}_1=2i\omega_0{\bf u}_0.
\ee
Therefore, in the parameter space the coordinates $(0,\pm \kappa_0,0)$ correspond to exceptional points (EPs), at which there
exist double eigenvalues with the Jordan chain, Fig.~\ref{fig3}.

In the vicinity of the EPs the expression $B=0$ for the stability boundary yields
\be{z7}
\Omega=\frac{4\beta_0\kappa\pm{\rm tr}{\bf D}\sqrt{\kappa^2-\kappa_0^2}}{4\kappa_0}\delta+
o(\delta).
\ee
Thus, in the $(\Omega,\delta)$-plane the domain of asymptotic stability is bounded in the first-order
approximation by the two straight lines \rf{z7}. When $\kappa$ goes
to $\pm\kappa_0$, the slopes of the both lines $\beta:=\Omega/\delta$ tend to the value $\beta=\pm\beta_0$, where $\beta_0$ is defined in \rf{cd1b}.
Extracting $\kappa$ from the equation \rf{z7} we find another representation for the stability boundary near exceptional points
\be{z8}
\kappa=\kappa_0\frac{4\beta\beta_0\pm{\rm tr}{\bf D}\sqrt{\beta^2-\beta_0^2+\left(\frac{{\rm tr}{\bf D}}{4}\right)^2}}
{4\beta_0^2-\left(\frac{{\rm tr}{\bf D}}{2}\right)^2}=\pm\kappa_0\left[1+8\left(\frac{\beta\mp\beta_0}{{\rm tr}{\bf D}}\right)^2\right]+
o\left((\beta\mp\beta_0)^2\right),
\ee
which has a canonical for the Whitney's umbrella singularity form $Z=X^2/Y^2$ \cite{HR95,Ki06,Ki07}.

Therefore, we explicitly demonstrated that the conical boundary of the domain of subcritical flutter for $A>0$ is structurally
unstable under the perturbation $\nu{\bf N}$.
With the increase of $\nu$ the cone opens up and simultaneously the plane $\delta=0$ foliates into two sheets
intersecting along the branch cuts $(\pm\infty,\pm\kappa_0]$ on the $\kappa$-axis, which are shown as thick lines in Fig.~\ref{fig3}.
The new surface has a couple of Whitney's umbrella singularities at the exceptional points $(0,\pm \kappa_0,0)$.
The domain of asymptotic stability, which for $\nu=0$ was adjacent to the conical domain
of subcritical flutter is now wrapped into the pockets of the two Whitney's umbrellas, selected by the inequality
$\delta{\rm tr}{\bf D}>0$.
With the increase in $\delta$ the stability boundary gradually tends to the conical surface with the flutter instability inside it,
Fig.~\ref{fig3}(a).

Inclusion of the non-conservative forces qualitatively changes the stability diagram
in the $(\Omega,\kappa)$-plane transforming the elliptic flutter domain of Fig.~\ref{fig2}(a) to
a larger one located between the hyperbolic branches, Fig.~\ref{fig3}(b). For $\nu=0$ and $\delta=0$ the $(\Omega,\kappa)$-plane
is stable, while for $\nu \ne 0$ and $\delta \rightarrow 0$ the stability domain dramatically shrinks to the branch cuts
$(\pm\infty,\pm\kappa_0]$. Consequently, under small perturbation $\nu{\bf N}$
a point in the $(\Omega,\kappa)$-plane, which was in the stability domain for $\nu=0$ can suddenly find itself
in the instability region when $\nu \ne 0$, similarly to the scenario described in the previous section.

For $A<0$ the conical stability boundary of Fig.~\ref{fig2}(c) unfolds into two surfaces with the Whitney's umbrella singularities
at the exceptional points $(0,\pm \kappa_0,0)$ as shown in Fig.~\ref{fig3}(c). The local approximations to the surfaces near the
singularities are given by the same equation \rf{z8}, where $\beta_0$ has a value different from the case when $A>0$.
For $\nu=0$ and $\delta\ne 0$ the stability domain in the $(\Omega,\kappa)$-plane is inside of the two hyperbolic regions
extended along the $\Omega$-axis, as shown in Fig.~\ref{fig2}(c). When $\nu \ne 0$ with the decrease of
$\delta$ the stability domain rotates around the origin until it is completely reoriented and shrunk into the branch
cuts $(\pm\infty,\pm\kappa_0]$ extended along the $\kappa$-axis, Fig.~\ref{fig3}(d). Due to such a reorientation one can again
observe sudden stabilization/destabilization at the very same values of $\Omega$ and $\kappa$ when $\nu$ is slightly changed.
Similar discontinuous change of planar stability diagrams was observed in \cite{HR95,Sh01,Sh04} for parametric oscillators coupled through the
gyroscopic and damping forces.

Therefore, the fundamental process of birth of a couple of exceptional points from a diabolical
point (whose location is marked by the asterisk in Fig.~\ref{fig3}), which has important consequences for
the wave propagation in chiral and dissipative media \cite{KMS05a,KMS05b,KKM03,BD03}, substantially determines
the stability of waves propagating
in rotating continua with frictional contact.

\section*{Conclusions}

The singing wine glass and the squealing brake are phenomena of the acoustics of friction,
which everyone encounters with almost every day. Despite of a seeming simplicity, their mathematical modeling is
not easy, and experiments with them are not satisfactorily reproducible. Several fundamental reasons are responsible
for this. First, the rotational symmetry of the continua makes the spectral mesh with the double eigenvalues at the nodes
generic for the unloaded bodies of revolution. Then, definite Krein signature at the eigenvalue crossings in the subcritical
rotation speed range requires dissipative and non-conservative perturbations to produce complex eigenvalues with the
positive real parts. The geometry of the domain of the dissipation-induced subcritical flutter is complicated by the
singularities related to multiple eigenvalues. The changes in the damping and stiffness matrices, which are unavoidable from one run
of the experiment to another one due to wear of the friction pads, provoke bifurcation of the flutter domain, which explains
why experimental results are not reproduced even after the accurate restoration of other well-tunable parameters.
Using perturbation theory of multiple eigenvalues we described typical bifurcations of the subcritical flutter
domain related to the birth of a couple of exceptional points from a diabolical point
due to non-Hamiltonian perturbation of a Hamiltonian system.

\section*{Acknowledgements}
The work has been supported by the research grant DFG HA 1060/43-1. The author expresses his gratitude to Professor E. Brommundt, Technische Universit\"at Braunschweig, for valuable discussions.
\appendix

\end{document}